\documentclass[11pt,a4paper]{article}
\pdfoutput=1
\usepackage{jheppub}

\usepackage{color} 
\usepackage{amsmath}
\usepackage{pifont}

\usepackage{verbatim}   
\usepackage{acronym} 
    
\usepackage{amsfonts}
\usepackage{amssymb}
\usepackage{mathrsfs} 
\usepackage{graphicx}
\usepackage{multirow} 
\usepackage{slashed} 
\usepackage{array}

\usepackage{graphicx}
\usepackage{epsfig}
\usepackage{lscape}
\usepackage{rotating}
\usepackage{epsf}
\usepackage{bm}
\usepackage{booktabs}
\usepackage{array}
\usepackage{caption}
\usepackage{subcaption}
\usepackage[utf8]{inputenc}
\usepackage{float}

\usepackage{multirow}
\usepackage{array,booktabs,colortbl,multirow}
\usepackage{colortbl,xcolor,colordvi,color}
\usepackage{rotating}
\allowdisplaybreaks

\newcommand{\beq}{\begin{align}}
\newcommand{\eeq}{\end{align}}
\def\be{\begin{equation}}
\def\ee{\end{equation}}
\def\bea{\begin{eqnarray}}
\def\eea{\end{eqnarray}}
\def\bitem{\begin{itemize}}
\def\eitem{\end{itemize}}
\newcommand{\bec}{\begin{center}}
\newcommand{\eec}{\end{center}}
\newcommand{\ba}{\begin{array}}
\newcommand{\ea}{\end{array}}

\newcommand{\chiV}[1]{(\overline{\chi} \gamma{#1} \chi)}
\newcommand{\chiA}[1]{(\overline{\chi} \gamma{#1} \gamma_5 \chi)}

\newcommand{\eV}[1]{(\overline{e} \gamma{#1} e)}
\newcommand{\eA}[1]{(\overline{e} \gamma{#1} \gamma_5 e)}

\newcommand{\muV}[1]{(\overline{\mu} \gamma{#1} \mu)}
\newcommand{\muA}[1]{(\overline{\mu} \gamma{#1} \gamma_5 \mu)}

\newcommand{\uV}[1]{(\overline{u} \gamma{#1} u)}
\newcommand{\uA}[1]{(\overline{u} \gamma{#1} \gamma_5 u)}

\newcommand{\dV}[1]{(\overline{d} \gamma{#1} d)}
\newcommand{\dA}[1]{(\overline{d} \gamma{#1} \gamma_5 d)}

\newcommand{\sV}[1]{(\overline{s} \gamma{#1} s)}
\newcommand{\sA}[1]{(\overline{s} \gamma{#1} \gamma_5 s)}

\newcommand{\cV}[1]{(\overline{c} \gamma{#1} c)}

\newcommand{\bV}[1]{(\overline{b} \gamma{#1} b)}
\newcommand{\bA}[1]{(\overline{b} \gamma{#1} \gamma_5 b)}

\begin{document}

\title{MeV Dark Matter: Model Independent Bounds} 

\affiliation{Instituto de F\'isica, Universidade de S\~ao Paulo, \\C.P. 66.318, 05315-970 S\~ao Paulo, Brazil}

\author{Enrico Bertuzzo,}
\author{Cristian J. Caniu Barros  and}
\author{Giovanni Grilli di Cortona}

\emailAdd{bertuzzo@if.usp.br, caniu@if.usp.br, ggrilli@if.usp.br}

\abstract{We use the framework of dark matter effective field theories to study the complementarity of bounds for a dark matter particle with mass in the MeV range. Taking properly into account the mixing between operators induced by the renormalization group running, we impose experimental constraints coming from the CMB, BBN, LHC, LEP, direct detection experiments and meson decays. In particular, we focus on the case of a vector coupling between the dark matter and the standard model fermions, and study to which extent future experiments can hope to probe regions of parameters space which are not already ruled out by current data.}

\maketitle

\section{Introduction and Statement of the Problem}

The nature of Dark Matter (DM) is one of the greatest puzzles of modern particle physics, as well as the nature of its interactions with the particles of the Standard Model (SM). Despite decades of experimental effort, the only interaction between DM and SM particles that has been confirmed experimentally is gravity. However, if other interactions are present, it would be desirable to have a model independent way to test how much parameter space has been actually tested and if there is room for discovery in future experiments. This can be achieved using Effective Field Theory (EFT) techniques, in which the only light degrees of freedom are the SM particles and the DM (see \cite{Cao:2009uw,Goodman:2010yf,Bai:2010hh,Goodman:2010ku} for some of the early works on the subject). The advantages of this approach are clear: it is as model independent as we can get, and it relies on just a few assumptions (namely, that the New Physics (NP) mediating the DM-SM interactions is heavier than the Electroweak (EW) scale and that it respects the SM gauge symmetry). On the other hand, since all the correlations between different operators (present in any concrete model) are lost, it is usually unfeasible to perform a global analysis involving more than a few operators. Still, many informations can be obtained, and it is on this framework that we will focus.

As is well known, most of the theoretical and experimental activity over the last decades has focused on the Weakly Interactive Massive Particle (WIMP) paradigm, {\it i.e.} a DM candidate with mass in the GeV-TeV range and with typical cross sections of Electroweak (EW) size. As a matter of fact, the region currently probed by Direct Detection (DD) experiments restrict to DM masses above $5$ GeV \cite{Akerib:2016vxi,Aprile:2017iyp}. In addition, one should also consider bounds from indirect detection and collider experiments, and possibly see if the simple thermal freeze-out mechanism can explain the observed DM abundance. We stress that a problem may arise in considering the LHC bounds applied to the EFT operators. Given the high centre of mass energy of the LHC, for any fixed cutoff around a few TeV a part of the produced events will have an energy above the cutoff. These events fall beyond the validity of the EFT, and as such should not be used in the computation of the bounds. This has motivated the use of simplified models \cite{DeSimone:2016fbz} as an useful intermediate step between the EFT and complete models. On the other hand, as shown in \cite{1502.04701}, it is possible to obtain robust collider limits if the centre of mass energy of the event is required to be below the EFT cutoff.~\footnote{See also \cite{Busoni:2013lha,Busoni:2014sya,Busoni:2014haa} for another prescription in the context of simplified-driven DM models.} One of the most surprising results of the analysis in \cite{1502.04701} is that, applying naive power counting to the Wilson coefficients and making a one-coupling-one-scale assumption (in the sense that only one cutoff scale $\Lambda$ and one coupling $g_*$ appear in all the EFT operators), only $g_* \gtrsim 2$ couplings are currently probed at the LHC. The same analysis has been applied to other operators in \cite{1607.02475,1607.02474}.

Given the plethora of null results challenging the WIMP paradigm, in the last few years the interest has turned to other regions of parameter space. In particular, the MeV region has emerged as an interesting possibility, with many well motivated models (see for example the SIMP case \cite{Hochberg:2014dra,Hochberg:2014kqa}, some models of asymmetric DM \cite{Falkowski:2011xh} and even some supersymmetric model \cite{Hooper:2008im,Feng:2008ya}). The purpose of this paper is to extend the model independent EFT analysis to the case of MeV DM, highlighting the complementarity of searches and pointing out to which extent and in which cases we should expect some signal in future experiments (especially in the $g_* =1$ case in which no LHC limits are available). Although less explored, some constraints are already available on the MeV DM parameter space. For instance, Cosmic Microwave Background (CMB) bounds already force the s-wave annihilation cross section into SM particles to be below the thermal one for masses below  $10$ GeV \cite{0905.0003,0906.1197,1109.6322,1106.1528,1604.02457}, and Big Bang Nucleosynthesis (BBN) bounds may put strong constraints on the annihilation into quarks \cite{Henning:2012rm}. This means that, if the DM has indeed a sub-GeV mass, the thermal freezeout paradigm has to be abandoned unless the dominant annihilation channels are p-wave suppressed. Moreover, bounds coming from colliders \cite{1103.0240,1607.02475,1607.02474}, meson decays \cite{1404.6599,1511.03728}, indirect searches \cite{1309.4091,1612.07698} and $Z$-physics at LEP \cite{Olive:2016xmw} must also be considered. Finally, the MeV region can in principle be probed in the future by DD experiments measuring DM-electron scattering \cite{1108.5383,1206.2644,1509.01598,Derenzo:2016fse,1608.02940,1703.00910,Cavoto:2017otc,Budnik:2017sbu} and in high intensity neutrino beam facilities \cite{Batell:2014yra,Coloma:2015pih,Frugiuele:2017zvx}. 

As can be seen, different SM particles are involved in the processes considered. As such, it looks like the only situation in which these constraints can be combined in a meaningful way to put bounds on the EFT coefficients is when the DM couples universally to all SM particles at the scale $\Lambda$ where the interactions are generated. However, this is not the only possibility. As shown in \cite{1402.1173,1411.3342,1605.04917}, dimension 6 operators mix in the running between $\Lambda$ and the low energy scale at which the experiments are performed. The result is that even if some operator is not present at high energy due to some unknown selection rule of the UV theory, it will be generated at low energy by the renormalization group running. Of course, how important this mixing is in imposing bounds depends crucially on the initial value of the Wilson coefficient at the scale $\Lambda$, and on the scale $\Lambda$ itself. Using this information, the complementarity of bounds in DM searches for a DM mass above $10$ GeV has been explored in \cite{1605.04917} for universal coupling to quarks, to leptons and to third generation fermions and in \cite{Alves:2016aa} for $Z'$ models. Moreover, bounds on pure leptophilic models coming from the LHC have been analyzed in \cite{1702.00016}. Models with the correct relic density for MeV dark matter are given, for example, in \cite{Dey:2016aa,Dedes:2017aa}. 

The paper is organized as follows. We first briefly recall the relevant operators which we will consider throughout the article. We then present current and future constraints that apply to MeV DM. Finally, we put together all the constraints taking properly into account the Renormalization Group Equations (RGE's) and show the available parameter space for some UV configuration.

\section{DM EFT and Running}\label{sec:EFT_summary}

As already mentioned in the Introduction, the main hypotheses behind DM-EFT are that the only light degrees of freedom below the cutoff are the SM particles and the DM, and that at the cutoff the whole SM gauge symmetry is respected. For definitiveness, we will always take the DM to be a Dirac singlet fermion $\chi$ with mass $m_{DM}$. At the scale $\Lambda$ the lagrangian is given by
\begin{equation}\label{eq:lagr}
	{\cal L} = {\cal L}_{SM}+ \overline{\chi} (i \slashed{\partial} - m_{DM}) \chi + \sum_i \left( \frac{g_*}{\Lambda} \right)^2 {\cal O}_{6i} + \dots\, ,
\end{equation}
where ${\cal L}_{SM}$ is the SM lagrangian and the dots represent all the operators constructed out of the SM particles only (see \cite{1008.4884} for the complete list). In writing Eq. (\ref{eq:lagr}) we are making a one-coupling-one-scale assumption (with the coupling $g_* = g_*(\Lambda)$ defined at the scale $\Lambda$) and we restrict the sum over dimension 6 operators only. For our purposes, this is justified since operators of dimension 5, do not mix under renormalization~\cite{1411.3342} (although they generate dimension 7 operators that can mix below the EW scale~\cite{1402.1173}). At the dimension 6 level, $32$ operators are present~\cite{1411.3342}. All of them can be written as the product of a DM and a SM current 
\begin{equation}
	O_{6i} = J^\mu_\chi J_\mu^{SM}\, ,
\end{equation}
where $J^\mu_\chi = \left\{\overline{\chi} \gamma^\mu \chi, \overline{\chi} \gamma^\mu \gamma_5\chi \right\}$ and the SM currents are given by
\begin{equation}
	J^\mu_{SM} = \left\{
	\begin{array}{lcl}
		\left( \overline{q}_L \gamma^\mu q_L, \overline{u}_R \gamma^\mu u_R, \overline{d}_R \gamma^\mu d_R, \overline{\ell}_L \gamma^\mu \ell_L, \overline{e}_R \gamma^\mu e_R , i H^\dag \stackrel{\leftrightarrow}{D}_\mu H \right) & \quad\quad & \mathrm{above}\,\, m_Z, \\[0.1cm]
		\left( \overline{u}\gamma^\mu u, \overline{d}\gamma^\mu d, \overline{e}\gamma^\mu e, \overline{u}\gamma^\mu \gamma_5 u,  \overline{d}\gamma^\mu \gamma_5 d, \overline{e}\gamma^\mu \gamma_5 e\right) & \quad\quad & \mathrm{below}\,\, m_Z,
	\end{array} \right.
\end{equation}
where the first line is appropriate in the unbroken EW phase (above $m_Z$) while the second one is appropriate in the broken EW phase (below $ m_Z$). Above $m_Z$ each operator appears three times, one for each generation (for simplicity, we will assume throughout the paper that the SM currents are flavor conserving), while below $m_Z$ the top quark does not appear, since it has being integrated out. Notice that since $i H^\dag \stackrel{\leftrightarrow}{D}_\mu H = \frac{\sqrt{g^2 + g'^2}}{2} (h+ v)^2 Z_\mu$, this operator does not appear below $m_Z$ because both the $Z$ and the $h$ bosons have been integrated out. 

The anomalous dimension matrices that mix the effective operators in the running above and below $m_Z$ have been computed in~\cite{1402.1173,1411.3342,1605.04917}, and are independent from the form of the DM current (since the DM current is a complete SM singlet, it does not contribute to the running). In order to compute the Wilson coefficient of the various operators at a scale $\mu$ relevant for the experiments we use the public code RunDM~\cite{RunDM}. A comment about the interpretation of the results is in order. Our exclusions are strictly valid for experiments performed with a typical energy $E \ll \Lambda$, since for these cases the mediator can obviously be integrated out. For experiments performed with $E \gtrsim \Lambda$ (as can be the case for LEP II or the LHC, as we will see below), the bounds should be computed keeping the mediator that generates Eq. (\ref{eq:lagr}) in the spectrum. However, such bounds are model dependent (the details of how the mediator couples to the DM are important when considering the resonant production). As shown in \cite{1502.04701}, the approach we will take gives a model independent bound even in the region in which the resonant production of the mediator is important, in the sense that we exclude a smaller region in parameter space. 

In what follows, we will focus on the so called D5 operator~\cite{Goodman:2010ku}, which is the product between the vector DM current and a vector SM current. In particular, we will analyze a {\it leptophobic} case, with universal coupling to quarks only,  ${\cal O}_{D5} = \sum_{i=1}^3 \left[  \bar{u^i}\gamma^\mu u^i + \bar{d^i} \gamma^\mu d^i \right] \bar{\chi} \gamma_\mu \chi $, and a {\it leptophilic} case with universal coupling to leptons only,  ${\cal O}_{D5} = \sum_{i=1}^3 \left[  \bar{\ell^i}\gamma^\mu \ell^i  \right] \bar{\chi} \gamma_\mu \chi $. We will briefly comment on other possibilities at the end of Section \ref{sec:summary}.

\section{Experimental bounds}\label{sec:Experimental_now}

\begin{table}
	\centering
	\begin{tabular}{c|c|c}
		Experiment & Process & Operators involved \\
		\hline
		\hline
		     \multirow{5}{*}{CMB}& \multirow{2}{*}{$ \overline{\chi} \chi \to e^+ e^-, \, \mu^+\mu^-$}&$\chiV{^\mu} \eV{_\mu}$, $\chiV{^\mu}\eA{_\mu}$, $\chiA{^\mu} \eA{_\mu}$\\
		    &&$\chiV{^\mu} \muV{_\mu}$, $\chiV{^\mu}\muA{_\mu}$, $\chiA{^\mu} \muA{_\mu}$\\ 
			& \multirow{3}{*}{$\overline{\chi} \chi \to \overline{M} M $}&$\chiV{^\mu} \uV{_\mu}$, $\chiA{^\mu}\uV{_\mu}$\\
		    &&$\chiV{^\mu} \dV{_\mu}$, $\chiA{^\mu}\dV{_\mu}$\\
			&&$\chiV{^\mu} \sV{_\mu}$, $\chiA{^\mu}\sV{_\mu}$\\
		    \hline
			\multirow{2}{*}{LEP} & \multirow{2}{*}{$e^+ e^- \to \gamma \overline{\chi} \chi$} & $\chiV{^\mu}\eV{_\mu}$, $\chiA{^\mu}\eV{_\mu}$ \\
								 & 															  & $\chiV{^\mu}\eA{_\mu}$, $\chiA{^\mu}\eA{_\mu}$ \\
		    \hline
			\multirow{4}{*}{LHC} & \multirow{4}{*}{$pp \to j \overline{\chi} \chi$} & $\chiV{^\mu}\uV{_\mu}$, $\chiV{^\mu}\dV{_\mu}$, $\chiV{^\mu}\sV{_\mu}$ \\
								 & 													& $\chiA{^\mu}\uV{_\mu}$, $\chiA{^\mu}\dV{_\mu}$, $\chiA{^\mu}\sV{_\mu}$ \\
								 & 													& $\chiV{^\mu}\uA{_\mu}$, $\chiV{^\mu}\dA{_\mu}$, $\chiV{^\mu}\sA{_\mu}$ \\
								 & 													& $\chiA{^\mu}\uA{_\mu}$, $\chiA{^\mu}\dA{_\mu}$, $\chiA{^\mu}\sA{_\mu}$ \\
		    \hline
			\multirow{3}{*}{Meson decays} & $\Upsilon \to \overline{\chi}\chi$ & $\chiV{^\mu}\bV{_\mu}$, $\chiA{^\mu}\bV{_\mu}$ \\
								 		  & $\Upsilon \to \gamma \overline{\chi}\chi$   & $\chiV{^\mu}\bA{_\mu}$, $\chiA{^\mu}\bA{_\mu}$ \\
										  & $J/\Psi \to \overline{\chi}\chi$  & $\chiV{^\mu}\cV{_\mu}$, $\chiA{^\mu}\cV{_\mu}$ \\
		    \hline
			\multirow{2}{*}{Direct detection} & $\chi e \to \chi e$ & $\chiV{^\mu}\eV{_\mu}$, $\chiA{^\mu}\eA{_\mu}$\\
											  & $\chi n \to \chi n$ & $\chiV{^\mu}\uV{_\mu}$, $\chiV{^\mu}\dV{_\mu}$,
	\end{tabular}
	\caption{\label{tab:current_constraints} Experimental constraints and operators probed. The $\overline{\chi}\chi \to \overline{M}M$ process refers to the DM annihilation into mesons (see Appendix for details). }
\end{table}

In this section we present the bounds from current or past experiments that can be applied to MeV DM. We summarize in Table~\ref{tab:current_constraints} all the experimental bounds and the operators to which they apply. 

\subsection{Bounds on the annihilation cross section}\label{ssec:CMB}

The DM annihilation cross section is bounded by CMB, BBN and indirect detection constraints. Self annihilation of dark matter particles may inject hadronic or electromagnetic energy in the intergalactic medium that may alter the thermal history of the Universe. Since recombination and primordial nucleosynthesis are well understood, bounds from the CMB and from BBN are in general important. 

In the case of CMB, free electrons remaining after recombination can scatter off CMB photons and modify the CMB power spectrum. CMB data from WMAP and Planck set limits on the annihilation parameter $P_{\mathrm{ann}}\equiv f(z) \langle \sigma v  \rangle/m_{DM}$, given in terms of the thermally averaged cross section $\langle \sigma v \rangle$ and the dark matter particle mass $m_{DM}$. The redshift dependent efficiency function $f(z)$ represents the amount of energy absorbed overall by the gas, and it is species dependent. The latest constraint from the Planck Collaboration is $P_{\mathrm{ann}}  < 4.1 \times 10^{-28}$ cm$^3$/s/GeV at $95\%$ C.L. ~\cite{1604.02457}. The CMB bound already rules out thermal s-wave annihilation cross sections for $m_{DM} \lesssim 10$ GeV~\cite{0905.0003,0906.1197,1109.6322,1106.1528,1604.02457}. In the future, Cosmic Variance Limited experiments have the potential to constrain $P_{ann} < 8.9 \times 10^{-29}$ cm$^3$/s/GeV~\cite{0905.0003}, {\it i.e.} a factor of $\sim 5$ more stringent than current bounds.

In the case of MeV DM, additional care must be taken in the computation of the CMB bound because the DM pair will annihilate to mesons rather than quarks. The coupling between mesons and the DM currents has been computed in~\cite{1611.00368} in the context of Chiral Perturbation Theory (see also Appendix~\ref{app:formulas} for more details), and in our computation we will consider all possible decays into light mesons and light leptons. We list in Table~\ref{tab:current_constraints} the operators involved.

In order to impose the bounds from CMB, we use the Equations in Appendix~\ref{app:formulas} taking the appropriate thermal average. We set $f(z)=1$ for the annihilation to mesons and we take the bound on the annihilation cross section to electrons from \cite{Slatyer:2015jla}. The choice $f(z)=1$ is an overestimate of the bound. It turns out, however, that even with $f(z)=1$ the meson contribution to $P_{ann}$ is always subdominant with respect to the electron one. Therefore, in setting the limits in Sec. \ref{sec:summary}, we will consider only the annihilation to electrons.

Turning to primordial nucleosynthesis, the injection of electromagnetic or hadronic energy in the intergalactic medium can dissociate already formed nuclei or can alter the neutron/proton ratio through pion exchange. The case of sub-GeV DM has been considered in Ref. \cite{Henning:2012rm}. Overproduction of $^3$He put bounds on the annihilation cross section into electrons, while deuterium overproduction put bounds on the $\chi\chi \to \bar{b}b$ annihilation cross section. The bound on $\chi\chi \to e^+ e^-$ is always weaker than the CMB bound, while for a DM mass between $4$ GeV and $20$ GeV the bound on $\langle\sigma v \rangle\bigl|_{\chi\chi \to b\bar{b}}$ is slightly stronger than the CMB one. As we are going to see, though, in the same region the bound coming from direct detection experiments is always stronger.

Concerning indirect detection searches, bounds on MeV DM coming from diffuse X-ray and Gamma ray observations have been computed in~\cite{1309.4091}, while more recently bounds from cosmic rays electrons and positrons have been computed in \cite{1612.07698} (see also \cite{Gonzalez-Morales:2017jkx}). In the decays of diffuse X-ray and Gamma ray, model independent bounds can be put on the annihilation cross section to electrons. For $m_{DM} \lesssim 30$ MeV, the limit coming from INTEGRAL e COMPTEL is of order $\langle \sigma v_{rel} \rangle \lesssim 10^{-27}$ cm$^3$/s, while for larger DM mass the bound becomes less and less stringent until it reaches the FERMI value $\langle \sigma v_{rel} \rangle \lesssim 10^{-24}$ cm$^3$/s for $m_{DM} \gtrsim 1$ GeV \cite{1309.4091}. Turning to cosmic ray data, limits can be extracted from Voyager 1 and AMS-02 \cite{1612.07698}. For masses around $m_{DM} \simeq 10$ MeV, the limits are slightly more stringent than those obtained from diffuse X and Gamma ray data. Still, they are roughly an order of magnitude weaker than those obtained from CMB. 

\subsection{Collider constraints: LEP and the LHC}\label{ssec:LEP}

As shown in~\cite{1103.0240}, mono-photon searches at LEP II can put bounds on the operators involving electrons listed in Table~\ref{tab:current_constraints}, although only the bounds on the operators $(\overline{\chi} \gamma^\mu \chi) (\overline{e}\gamma_\mu e)$ and $(\overline{\chi} \gamma^\mu \gamma_5 \chi) (\overline{e}\gamma_\mu \gamma_5 e)$ have been computed. For $m_{DM} \lesssim 20$ GeV, the constraint on the two operators is the same and is as strong as $\Lambda \sim 500$ GeV for Wilson coefficients equal to $1$. As explained in the introduction, to ensure the validity of the EFT in the considered events, the cut $E_{\rm cm} = \sqrt{(p_{DM,1} + p_{DM,2} + p_\gamma)^2} < \Lambda$ should be imposed, analogous to what proposed in~\cite{1502.04701,1607.02475,1607.02474}. A complication arises, however. The monophoton data were collected with centre of mass energies scanning between $180$ GeV and $209$ GeV, so that it is not completely well defined which energy scale should be used in the computation of the Wilson coefficient. Since the analysis of Ref.~\cite{1103.0240} was performed supposing $E_{\rm cm} = 200$ GeV, in what follows we will simply take all the coefficients computed at a scale $\mu \simeq 200$ GeV, and declare that the scales below this energy cannot be probed inside the validity of the EFT. 

Other signatures can be better exploited at the LHC. In particular, the strongest experimental constraints come from mono-jet searches \cite{Khachatryan:2014rra,Aad:2015zva,Aaboud:2016tnv,Sirunyan:2017hci} can be used to put bounds on the operators listed in Table~\ref{tab:current_constraints}. We recast the ATLAS search \cite{Aad:2015zva}, imposing the cut $E_{\mathrm{cm}}<~\Lambda$, where $E_{\mathrm{cm}}$ is the centre of  mass energy of the process, $E_{\mathrm{cm}}\equiv \sqrt{(p_{DM,1} + p_{DM,2} + p_j)^2}$. 
The ATLAS analysis taken in consideration allows for multiple jets and the cuts require at least one jet with a $p_T>120$ GeV, allowing for the presence of soft and collinear jets. 
We implement the dimension six operators in Feynrules \cite{Alloul:2013bka} and use MadGraph5\_aMC@NLO \cite{Alwall:2014hca} to generate events at matrix element level with the mono-jet topology. 
We then pass the events to PYTHIA 6 for parton showering and hadronisation \cite{Sjostrand:2006za}. In particular, we generate 200k events at parton level with 0-, 1- and 2-jets and we perform the final recast with MadAnalysis5 \cite{Conte:2012fm}, modifying existing code \cite{1607.02475,Bruggisser:github}. 
One of the main outcomes of the cut is that for couplings $g_* = 1$ no bound is found, while for $g_* = 4 \pi$ we find that, for $m_{DM} \lesssim 100$ GeV, the region $400$ GeV $\lesssim \Lambda \lesssim 12$ TeV is excluded. In particular, the region $\Lambda \lesssim 400$ GeV is not currently probed by the LHC, not even for large couplings. 

Let us conclude with some remarks on the bounds that can be extracted from $Z$ physics. When the ``Higgs portal'' operators $(\overline{\chi} \gamma^\mu \chi) (i H^\dag \stackrel{\leftrightarrow}{D}_\mu H)$ and $(\overline{\chi} \gamma^\mu \gamma_5 \chi) (i H^\dag \stackrel{\leftrightarrow}{D}_\mu H)$ are generated at the scale $\Lambda$, they have two effects. First, they contribute with a threshold correction to the evolution of the four fermion operators (through the SM coupling between the Z boson and the SM fermions) \cite{1411.3342}. Second, they generate a $Z-\chi-\chi$ interaction that can be bound from the $Z$ invisible width, $\Gamma_{inv}^{NP} < 1.5$ MeV~\cite{Olive:2016xmw}. Both effects can be used to set stringent limits on the parameter space.

\subsection{Mesons decays}\label{ssec:mesons}

For MeV DM the invisible decays of mesons play an important role. In what follows, we will focus for simplicity only on the invisible decays arising at tree level. In particular, we consider the decays $\Upsilon \to \overline{\chi}\chi$ and $J/\psi \to \overline{\chi}\chi$~\cite{1404.6599}.\footnote{The decay $\Upsilon \to \gamma \overline{\chi}\chi$ is relevant for vector axial quark  currents and constrain $\Lambda$ to be above $\mathcal{O}(50)$ GeV~\cite{1511.03728}.} 
Since $m_\Upsilon \simeq 10$ GeV and $m_{J/\psi} \simeq 3$ GeV, the bounds will be relevant for $m_{DM} < 5$ GeV and $m_{DM} < 1.5$ GeV, respectively, and the probed $\Lambda$ scales will be above the $\Upsilon$ and the $J/\psi$ masses. The meson masses are also the typical energy of the process, at which the relevant Wilson coefficients must be computed. The angular momentum and C/P transformation properties of the initial state determine which operators is involved in the decay process, and the different possibilities are listed in Table~\ref{tab:current_constraints}. 

The $90\%$ C.L. constraint on the branching ratio for invisible decays of $\Upsilon(1S)$ and $J/\Psi$ measured by BABAR and BES are
$BR(\Upsilon(1S)\to\mathrm{invisible})<3.0\times 10^{-4}$ and $BR(J/\Psi\to\mathrm{invisible})<7.2\times 10^{-4}$. On the other hand, the meson decays to $\bar{\nu}\nu$ via a $Z$ boson are negligible: $BR(\Upsilon(1S)\to \bar{\nu}\nu)<9.8\times 10^{-6}$ and $BR(J/\Psi\to \bar{\nu}\nu)<2.77\times 10^{-8}$. It is therefore enough to compute the branching fraction for the bound state to decay to DM.
For each process, the bounds are practically equal for all the operators involved, and can be as strong as $\Lambda \gtrsim 200$ GeV for the $\Upsilon \to \overline{\chi}\chi$ decay for an UV Wilson coefficient of order unity~\cite{1404.6599,1511.03728}. Although {\it per se} the bound is not very strong (much weaker than the LEP or LHC one, for instance), it is helpful to close in a model independent way the small $\Lambda$ window left open by colliders. 
In the future, according to~\cite{BelleII}, we can expect roughly a factor of 10 improvement in the sensitivity of $BR(\Upsilon \to \chi\chi)$ at Belle II, which translates in an improvement of the bound on $\Lambda$ of about a factor of $2$.

\subsection{Direct Detection Experiments}\label{ssec:DDcurrent}

Direct detection experiments have set stringent constraints on the dark matter nucleon scattering cross section for dark matter masses larger than $\sim 5$ -- $10$ GeV. Indeed, for spin independent scattering LUX and Xenon1T have reached a cross section limit of $\sim 10^{-46}$~cm$^2$ for $m_{DM}\sim 30$ GeV at $90\%$ C.L.  \cite{Akerib:2016vxi,Aprile:2017iyp}. On the other hand, the low mass region is weakly probed. Indeed for light dark matter, the fraction of initial energy transferred to the nucleus is suppressed by $m_{DM}/m_N$, leading to negligible recoil energy. The LUX and Xenon1T experiments probe DM scattering with nucleons only down to masses of about $5$ GeV, the maximum exclusion being $\sigma_{SI} \lesssim 10^{-42}\, {\rm cm}^2$. However, the forecasted sensitivity of SuperCDMS in the Si and Ge modes will be able to probe the MeV parameter region down to $m_{DM} \simeq 400$ MeV, with sensitivity to exclude DM-nucleon cross sections down to $10^{-(39 \div 43)}\, {\rm cm}^2$ (depending on the DM mass)~\cite{1610.00006}.

Although the recoil energy for a light dark matter particle scattering off nuclei is negligible, the kinetic energy involved in the process is large enough to ionize the target atom. Experiments like Xenon10 and Xenon100 could detect the ionization of a single atom \cite{1206.2644}. The Xenon10 experiment, using only 12 days of  calibration data,  can weakly probe the scattering cross section of MeV dark matter on free electrons as low as $\sim10^{-38}$ cm$^2$. Despite the weak bound, future experiments (or the analysis of data of current experiments as Xenon100 and LUX) could produce competitive limits. Moreover, different materials and process can increase the limit on DM scattering off electrons~\cite{1108.5383,1509.01598,Derenzo:2016fse,1608.02940,1703.00910,Cavoto:2017otc,Budnik:2017sbu}. The most promising process seems to be the DM scattering off electrons in semiconductor targets~\cite{1509.01598}, which can reach a sensitivity of about $\sigma_e \simeq 10^{-(43 \div 42)}$ cm$^2$. We refer to Table~\ref{tab:current_constraints} for the list of the operators contributing to the DM-electron scattering and that give an unsuppressed contribution to Spin Independent (SI) direct searches. The Wilson coefficients should be computed at the scale $\mu \simeq 1$ GeV.~\footnote{Strictly speaking, the typical energy scale for DM scattering off electrons would be below the GeV. However, given the difficulties in the computation of the running of the Wilson coefficients in the regime where QCD is non perturbative, we will consider a typical scale $\mu \simeq 1$ GeV also for DM-electron scattering.}

\subsection{Relic Density}

Now we discuss how to obtain the correct relic density for MeV DM. As pointed out in Section~\ref{ssec:CMB}, CMB bounds generically rule out a thermal annihilation s-wave cross section for DM masses $m_{DM} \lesssim 10$ GeV. This leaves open the possibility that either the DM is produced thermally via p-wave annihilation, or that a non-thermal production mechanism must be invoked.

In the case of the D5 operator (see the end of Sec. \ref{sec:EFT_summary}), the annihilation cross section is s-wave and the relic abundance should be produced non-thermally. Following Ref. \cite{Chu:2013jja}, we can compute the relic abundance inside the validity of the EFT if we suppose that the reheating temperature at the end of inflation is small enough not to produce the degrees of freedom that have been integrated out, {\it i.e.} $T_{RH} < \Lambda$. In this case, most of the DM production happens at temperatures much larger than the mass of DM or SM particles. Considering for simplicity a universal coupling to all fermions, we get \cite{Chu:2013jja}
\be
\Omega_{DM} h^2 \simeq \frac{m_{DM} Y_\chi^0}{3.6 \times 10^{-9} {\rm GeV}} \simeq \frac{m_{DM}}{3.6 \times 10^{-9} {\rm GeV}} \frac{4}{3} \frac{384}{(2\pi)^7} \left( \frac{45}{\pi g_*^s} \right)^{3/2} \frac{g_*^4 M_{PL}}{\Lambda^4} \left( T_{RH}^3 - T_0^3 \right)\, ,
\ee
where $T_{RH}$ is the reheating temperature, $T_0 = 2.7$ K the present temperature and $g_*^s$ the number of effective degrees of freedom in entropy. Imposing $\Omega_{DM}h^2 \simeq 0.12$, we get that the value of $\Lambda$ able to reproduce the observed relic abundance is
\be\label{eq:correct_relic}
\Lambda \sim g_* \left( 6 \times 10^7 \, {\rm GeV} \right) \left( \frac{m_{DM}}{0.01\, {\rm GeV}} \right)^{1/4} \left( \frac{T_{RH}}{1000\, {\rm GeV}} \right)^{3/4}\, .
\ee
As we are going to see in Section \ref{sec:summary}, this region of parameter space is not currently probed, and will not be probed in future experiments.

\section{Summary of the constraints}\label{sec:summary}

\begin{figure}[tb]
	\centering
	\includegraphics[width=.49\textwidth]{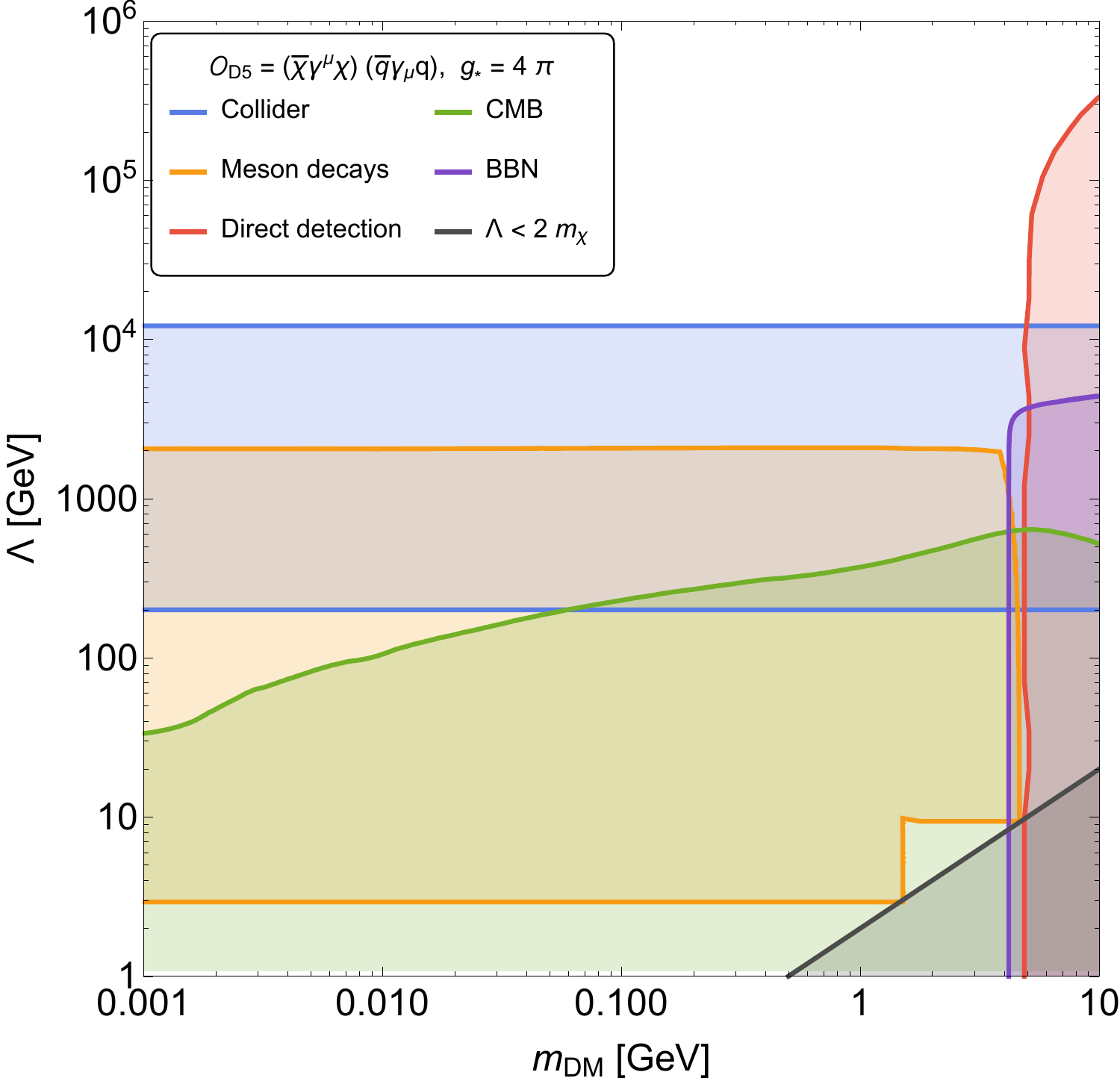} 
	\includegraphics[width=.49\textwidth]{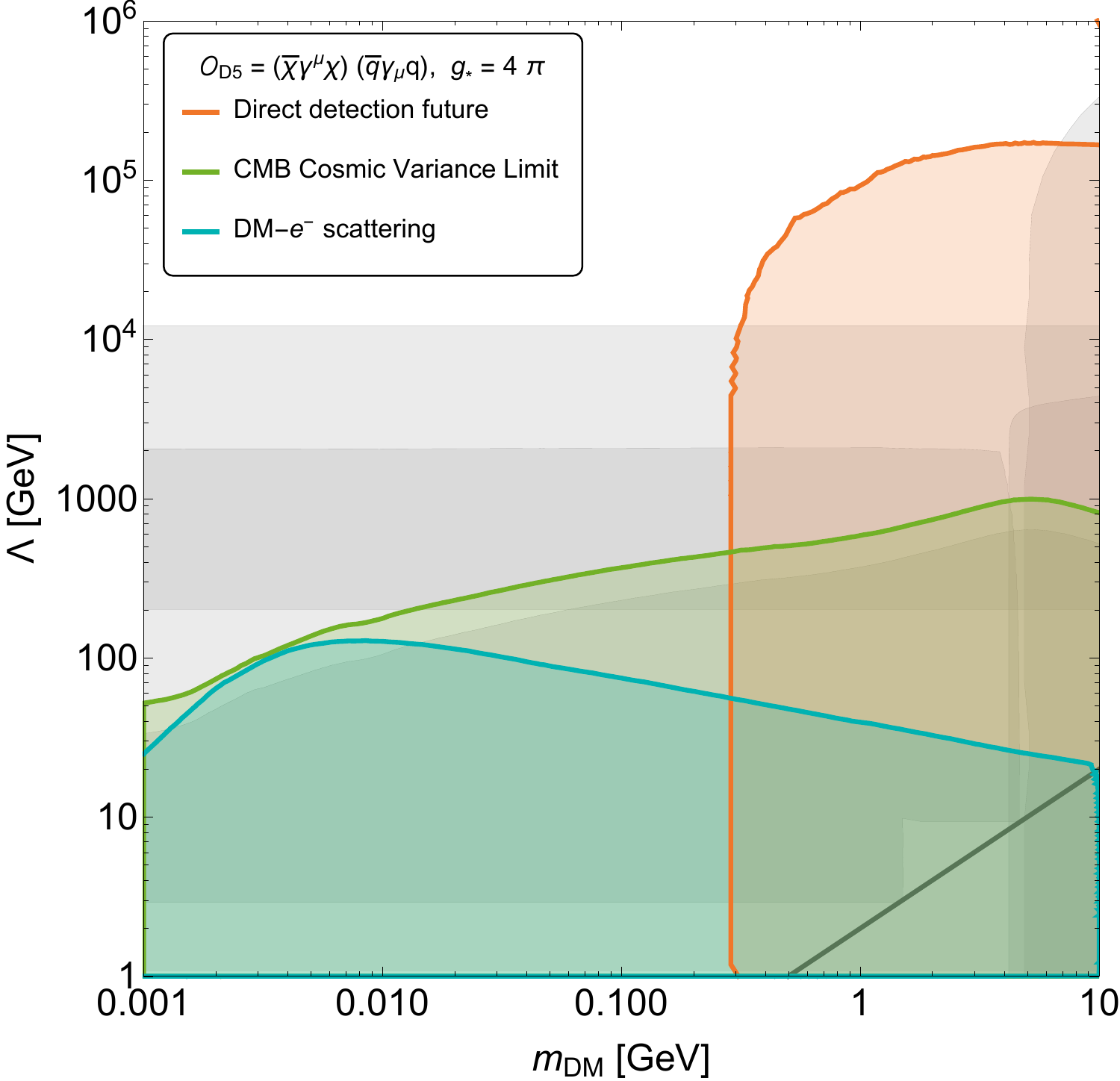}\\
	\includegraphics[width=.49\textwidth]{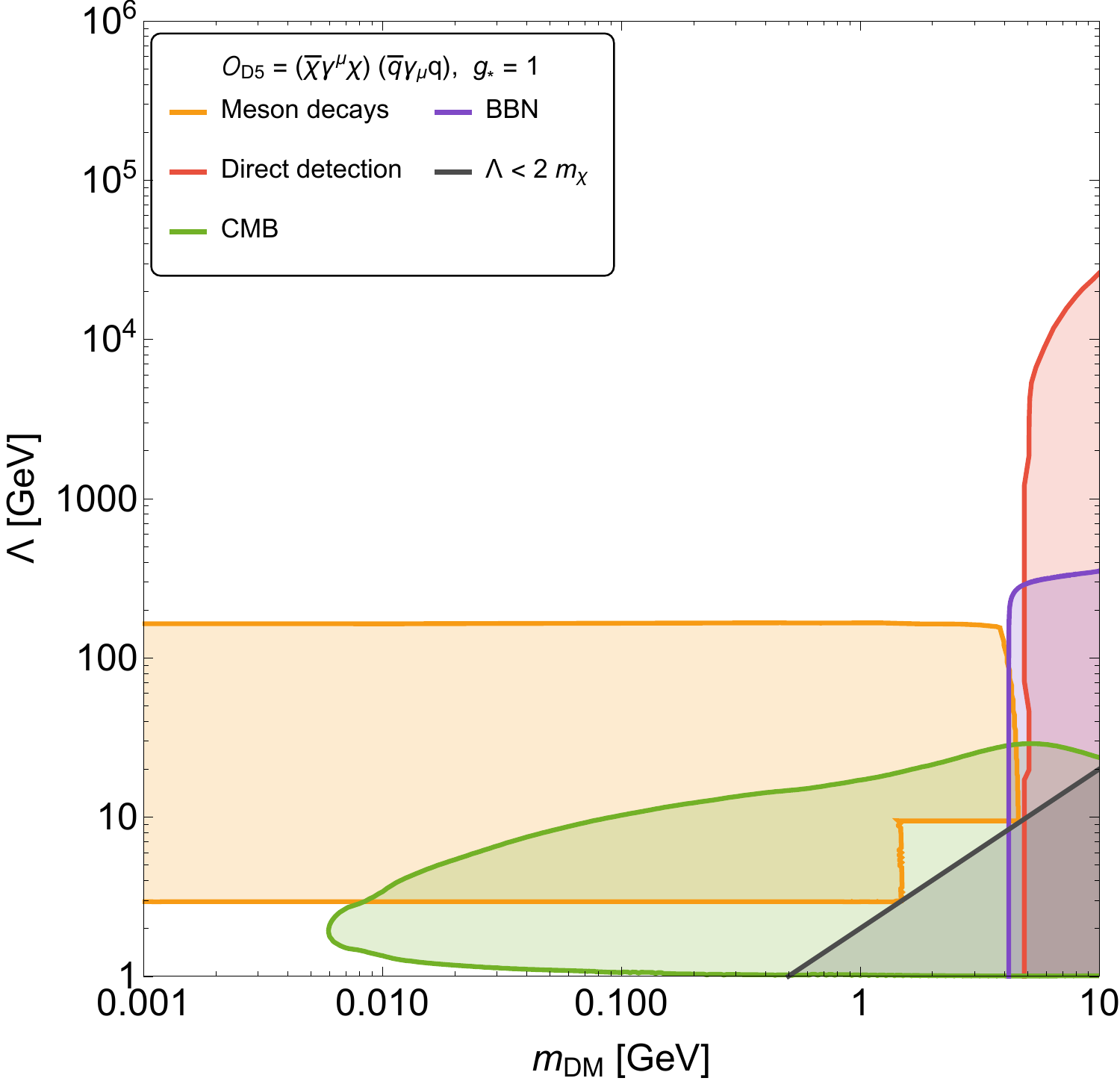} 
	\includegraphics[width=.49\textwidth]{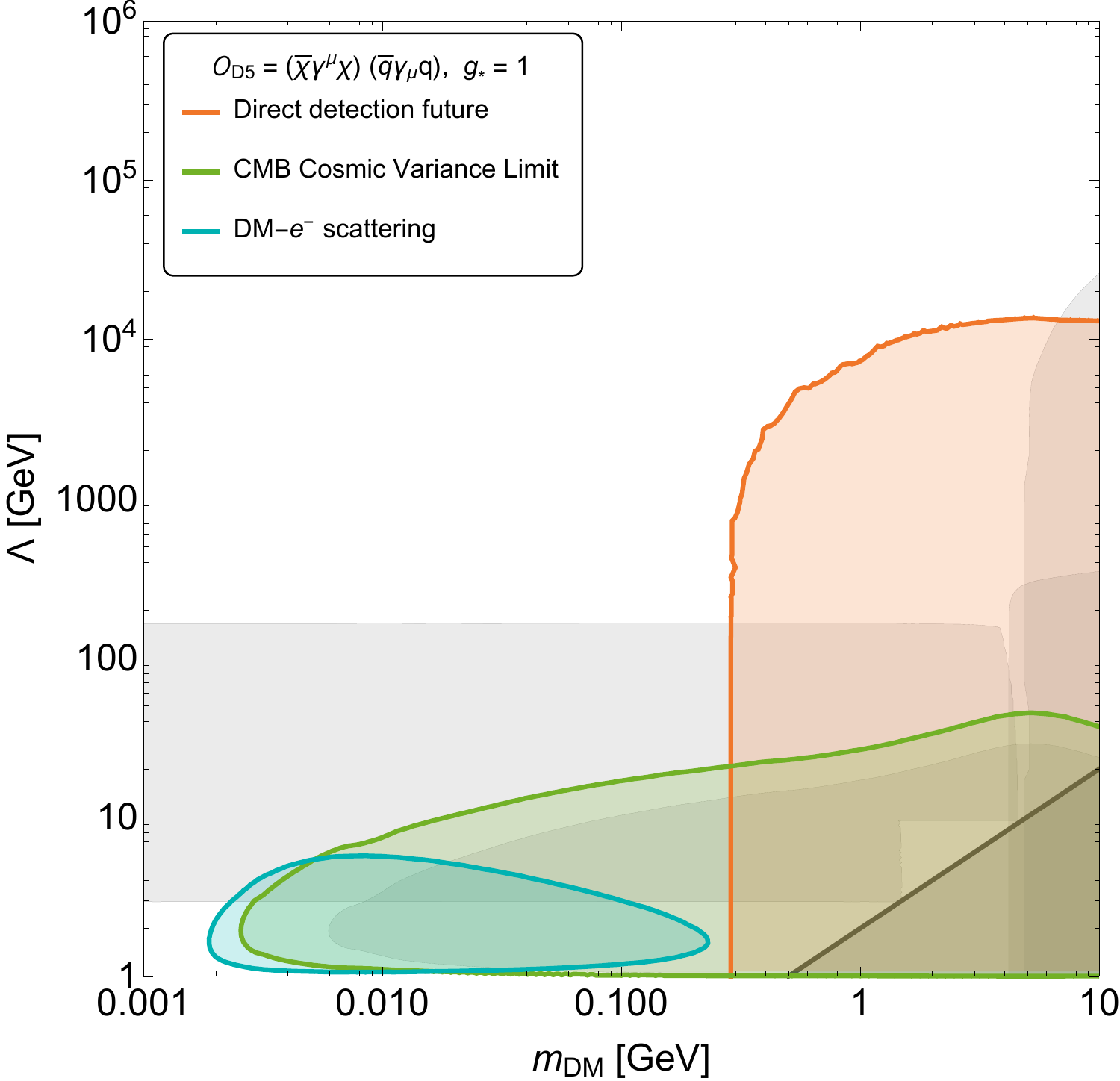}
\caption{Experimentally excluded region in the $(m_{\mathrm{DM}},\,\Lambda)$ parameter space for the case of SM singlet Dirac fermion dark matter with flavour universal couplings with quarks. In the left panels the blue, red, yellow, green and purple regions are ruled out respectively by collider (LEP and LHC), direct detection (LUX), meson decays (BaBar and BES), CMB experiments and BBN. The grey region represents the limit of validity of the EFT, $\Lambda < 2\, m_{\mathrm{DM}}$. In the right panel the green, emerald and orange regions will be probed respectively by future CMB, DM electron scattering and direct detection (SuperCDMS) experiments, while the grey area is already excluded by current experiments. The two upper panels consider an effective coupling $g_*=4\pi$ for current (left) and future (right) experiments. The lower panel shows results for an effective coupling $g_*=1$.  \label{fig:QU}}
\end{figure}

In this section we compare all the present bounds and future sensitivities discussed in Section~\ref{sec:Experimental_now}. We are interested in determining in which cases, if any, the parameter space to be probed by future experiments is already ruled out in a model independent way by current results. In the dark matter effective field theory we have two mass scales, the dark matter mass and the cut off scale $\Lambda$, and one coupling. We will present our results in a two-dimensional parameter space ($m_{\mathrm{DM}},\,\Lambda$), fixing the effective coupling to the maximum value allowed by perturbativity, $g_* = 4\pi$, and to $g_* = 1$. For concreteness, and to avoid bounds from structure formation \cite{Bond:1983hb,Menci:2016eui}, we focus on the mass range $1$ MeV $\leq m_{DM} \leq$ 10 GeV.~\footnote{See for instance Ref. \cite{Fichet:2017bng} for bounds on keV DM.} We consider two benchmark models for the operator $\mathcal{O}_{D5}$ introduced at the end of Section \ref{sec:EFT_summary}: universal couplings to all quarks (Sec.~\ref{subsec:UQ}) and universal couplings to all leptons (Sec.~\ref{subsec:UL}). We will comment on other possibilities in Section \ref{subsec:UF}.

\subsection{Universal couplings to quarks: leptophobic case}\label{subsec:UQ}

We start considering the case in which the DM vector current couples only to the quark vector current with flavor universal couplings. At the scale $\Lambda > m_Z$ the effective interactions are described by
\be\label{eq:D5quarks}
\mathcal{L}\supset \frac{g_*^2(\Lambda)}{\Lambda^2} \sum_{i=1}^3 \left[  \bar{u^i}\gamma^\mu u^i + \bar{d^i} \gamma^\mu d^i \right] \bar{\chi} \gamma_\mu \chi,
\ee
while for $\Lambda < m_Z$ the top has to be consistently integrated out.
We would therefore expect only experiments involving interactions between quarks and dark matter to contribute. However the running of the Wilson coefficient will induce also low energy couplings with leptons $c_{V}^{(\ell)}$. 
Solving the RGE's in the leading log approximation \cite{1605.04917,1702.00016}, we get
\be
c_{V}^{(\ell)}(1\,\mathrm{GeV}) \simeq \frac{4 \alpha_{em}}{3 \pi} g_*^2(\Lambda) \left[\theta(\Lambda-m_Z) \ln \frac{\Lambda}{m_Z} +\frac{1}{2}\ln \frac{\Lambda}{\mathrm{GeV}}  \right],
\ee
where the Heaviside function is needed when $\Lambda$ is below the EW scale.
This leads to the possibility to get limits on $\Lambda$ from LEP and from future DM-electron scattering experiments. 

In Fig. \ref{fig:QU} we show the excluded parameter space in the ($m_{DM},\,\Lambda$) plane. In the upper (lower) panels we show the results for $g_*=4\pi$ ($g_*=1$), while the left (right) panels show the current (future) exclusions. Let us start with $g_* = 4\pi$. In the upper left panel, the large couplings at the scale $\Lambda$ lead to important LHC bounds (blue region), as discussed in Sec. \ref{ssec:LEP}. Moreover, the induced coupling with electrons is also sizeable, such that the limits from LEP also apply (this is the reason why the collider bounds goes down to $\Lambda \simeq 200$ GeV). The yellow area is excluded by mesons decay. In particular, the upper limit of about $2$ TeV is set by the $\Upsilon(1s)$ to invisible decay (hence it applies to DM masses up to $5$ GeV). The lower limit is instead set by a combination of the bounds of the $\Upsilon(1s)$ and $J/\Psi$ decays. The $J/\Psi$ decay sets a stronger lower limit for DM masses up to $1.5$ GeV, where we clearly see the threshold due to the closure of this channel. The limits from CMB are able to cover the whole range of $\Lambda$ not covered by collider and meson decays, since the annihilation cross section is s-wave (see Appendix \ref{app:formulas}). As expected, the direct detection bounds from LUX are relevant only for $m_{DM}\gtrsim5$ GeV. Concerning future experiments (upper right panel), DM-electron scattering and limits from CMB will only be able to probe a large part of the parameter space already ruled out. Interesting information will instead come from future direct detection experiments as Super-CDMS. 

Turning to $g_* = 1$, as expected the bounds are much less severe. First, there are no LHC bounds. This is due to the issue of the validity of the EFT, as discussed in Section \ref{ssec:LEP}. In addition, the $g_*$ coupling at $\Lambda$ is now too small to induce a relevant coupling to electrons, in such a way that also the LEP bound is not present. The limits from mesons decay and from LUX are weaker because they just rescale with the coupling. As for the CMB limits, we see that for DM masses above $0.01$ GeV, the bound is basically a rescaling of the bound in the upper panel (although, being the coupling generated through running, there are some distortions). Below this mass we suddenly lose sensitivity due to the fact that we compute the Wilson coefficient at a scale $\mu \simeq 1$ GeV, instead of $\mu \simeq 2 m_{DM}$. The same happens in the right panel for the region probed by DM scattering off electrons. Future direct detection experiments as SuperCDMS will set strong limits on the scale $\Lambda$ for DM masses above $\sim 300$ MeV. For small couplings competitive bounds may come from DM-electron experiments for the region with small DM mass and small $\Lambda$.

Let us stress that, comparing the bounds in Fig. \ref{fig:QU} with Eq. (\ref{eq:correct_relic}), we see that not even the future experiments will be able to probe the region in which non-thermal relic production is effective.

\begin{figure}[tb]
	\centering
	\includegraphics[width=.49\textwidth]{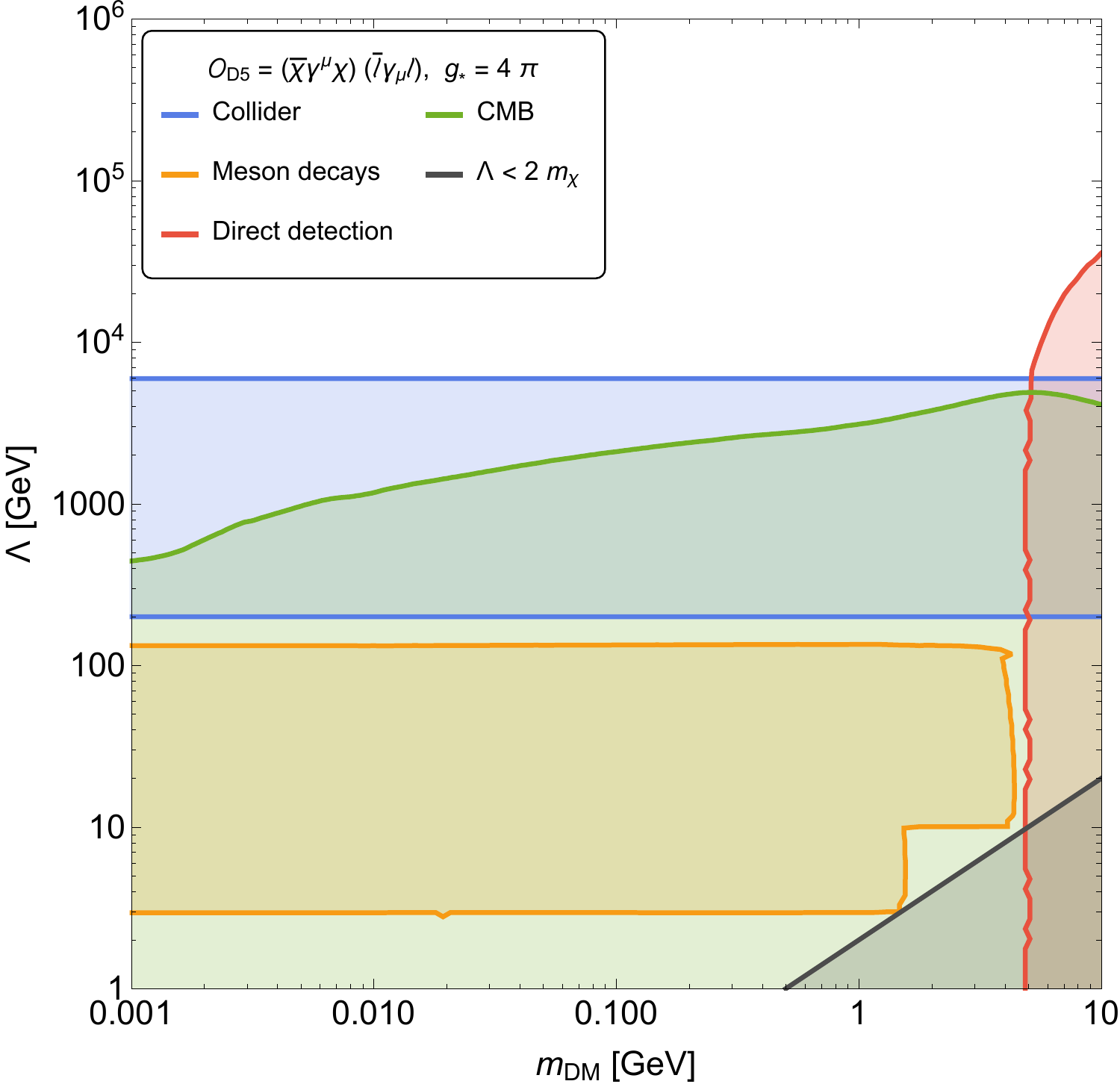} 
	\includegraphics[width=.49\textwidth]{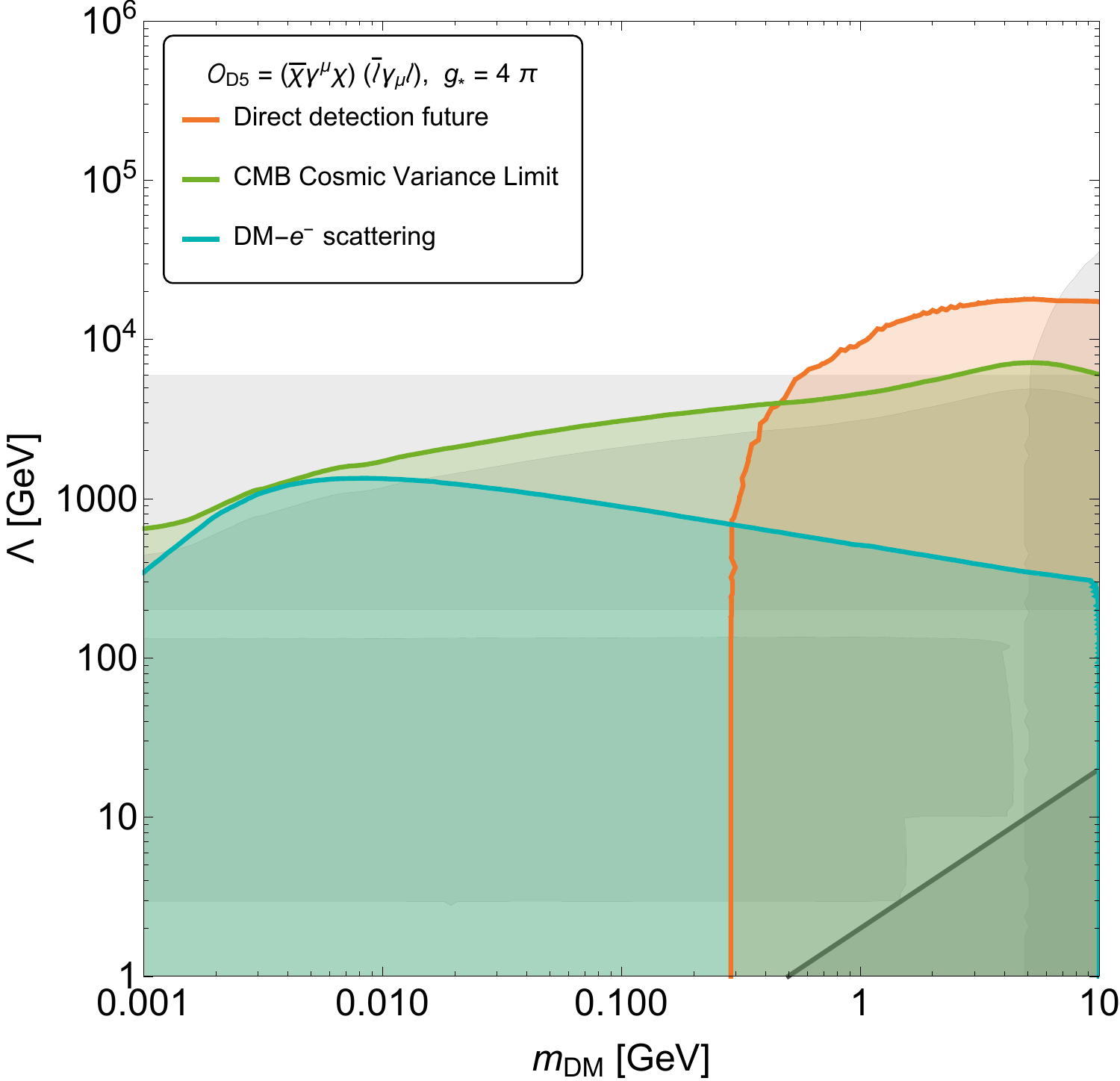}\\
	\includegraphics[width=.49\textwidth]{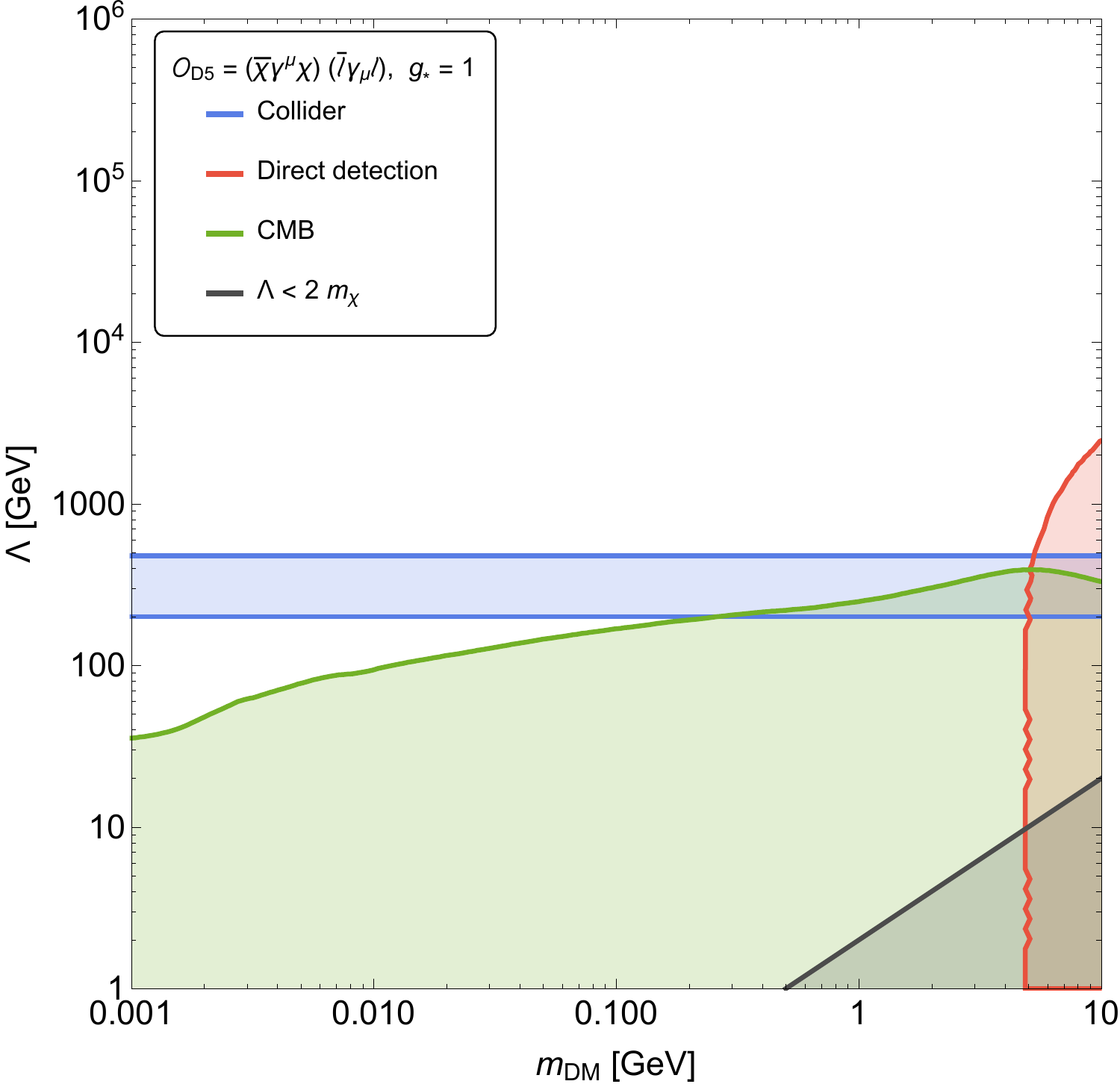} 
	\includegraphics[width=.49\textwidth]{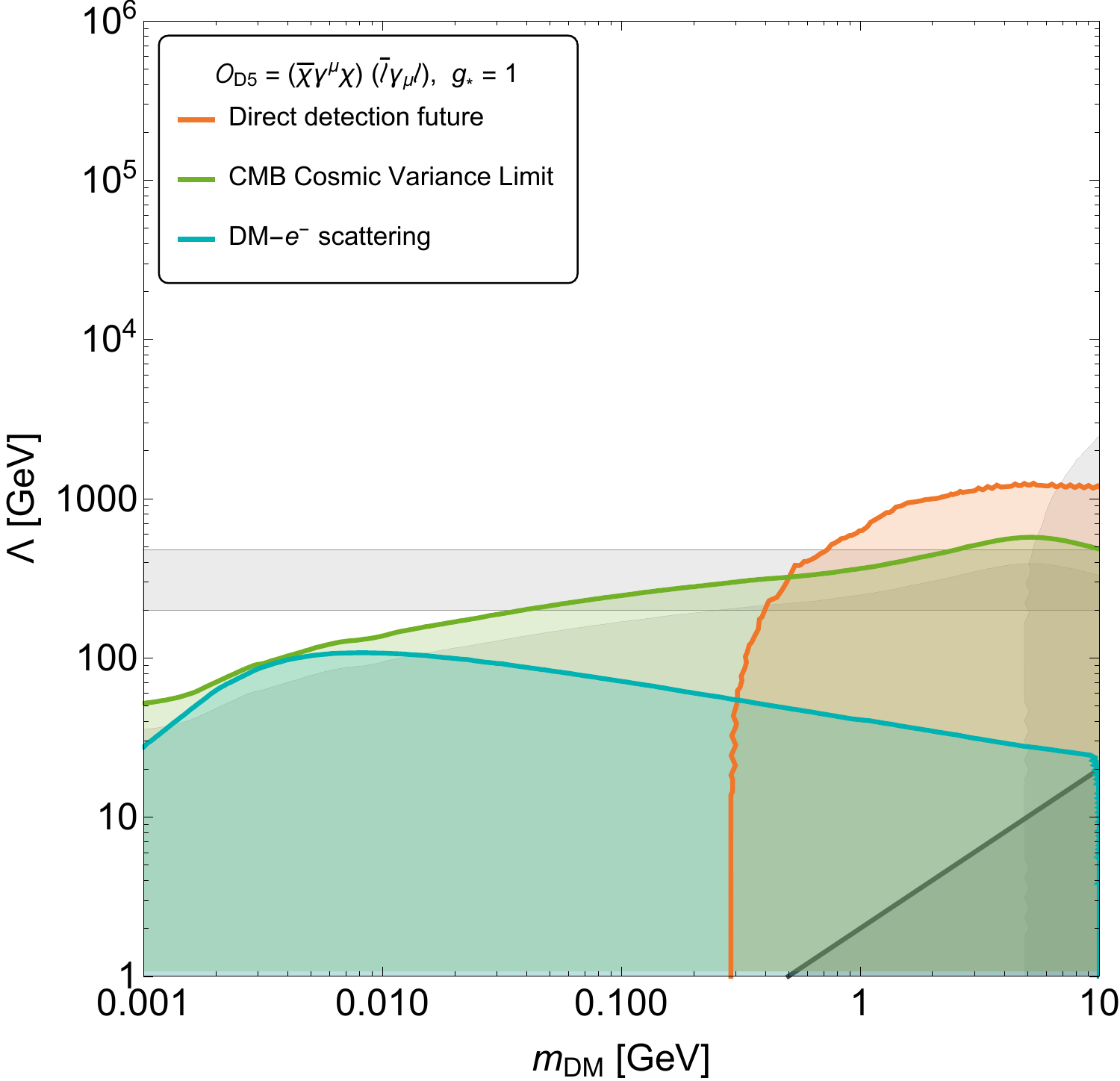}
\caption{Same as Fig. \ref{fig:QU}, but for universal couplings to leptons.\label{fig:LU}}
\end{figure}
\subsection{Universal couplings to leptons: leptophilic case}\label{subsec:UL}

As a second scenario, we consider a SM gauge invariant effective field theory where the dark matter current couples universally only to leptons:

\be
\mathcal{L}\supset \frac{g_*^2(\Lambda)}{\Lambda^2} \sum_{i=1}^3 \left[  \bar{\ell^i}\gamma^\mu \ell^i \right] \bar{\chi} \gamma_\mu \chi.
\ee
In this case, the running induces low energy Wilson coefficients with light quarks \cite{1605.04917}
\bea
c_V^{(u)}(1\,\mathrm{GeV}) &\simeq& \frac{4 \alpha_{em}}{3 \pi} g_*^2(\Lambda) \ln \frac{\Lambda}{\mathrm{GeV}}  \nonumber\\
c_V^{(d)}(1\,\mathrm{GeV}) &\simeq&  -\frac{2 \alpha_{em}}{3 \pi} g_*^2(\Lambda) \ln \frac{\Lambda}{\mathrm{GeV}} .
\eea
The presence of this couplings makes possible constraints from direct detection experiments. Indeed, the contact interaction at the scale $\Lambda$ do not involve light quarks and the dark matter nucleon scattering cross section comes only from radiatively induced interactions with light quarks. The same happens for meson decays. This result is visible in the top panels of Fig. \ref{fig:LU}, where we show the constraints for this scenario with $g_*(\Lambda)=4\pi$. The strongest limits comes from colliders (blue region), via the LEP experiment, that exclude $\Lambda$ to be between $\sim200$ GeV and $\sim6$ TeV and from CMB (green), that strongly constrains the annihilation cross section to electrons. The constraints form meson decays (yellow) are weaker, due to the fact that the couplings to light quarks arise only radiatively, and exclude $\Lambda$ between $\sim3$ and $\sim100$ GeV, for dark matter masses below $5$ GeV. For a dark matter heavier than $5$ GeV, the strongest limits are due to LUX (red). The right panel of the first row shows the reach of future CMB (green), DM-electron scattering (emerald) and direct detection experiments (orange).

The bottom panels show the exclusions for a coupling $g_* = 1$. With such small couplings, the running of the Wilson coefficients is not enough to set bounds from meson decays and the bounds from LUX are reduced. The absence of the meson decay limits leaves unexplored a small region between the LEP lower limit and the CMB bound. This region will be hardly covered with the next generation CMB or DM-electron scattering experiments.

As it happens in the leptophobic case, comparing the bounds in Fig. \ref{fig:LU} with Eq. (\ref{eq:correct_relic}), we see that also for the leptophilic case the region in which non-thermal relic production produces the correct relic abundance is not and will not be probed by future experiments.

\subsection{Other cases}\label{subsec:UF}

\noindent
Here we discuss a few other possibilities that may arise. First, we consider the situation in which the D5 operator involves a universal coupling with all the SM fermions. Since in this case all the couplings are turned on at tree level, the running will have a rather minor effect on the bounds. In fact, we have checked that the excluded regions correspond to the strongest constraints coming from the leptophobic and leptophilic case analyzed in the two previous sections: for $g_* = 4\pi$, all the region below $\Lambda \lesssim 10$ TeV is probed, with the bound set by the LHC limit. On the other hand, for $g_*=1$ the upper bound is dominated by the LEP constraint, with $\Lambda \lesssim 500$ GeV excluded.

Another interesting situation is given by the so called ``Higgs portal''. In this case, only one between the operators $(\overline{\chi} \gamma^\mu \chi) (i H^\dag \stackrel{\leftrightarrow}{D}_\mu H)$ and $(\overline{\chi} \gamma^\mu \gamma_5 \chi) (i H^\dag \stackrel{\leftrightarrow}{D}_\mu H)$ is turned on at the scale $\Lambda$. The coupling to fermions arise below $m_Z$, once the $Z$ boson is integrated out. As discussed in Section \ref{ssec:LEP}, the Higgs portal operators induce a $Z\chi\chi$ coupling that can be bound by the $Z$ invisible decay width. This bound turns out to be strong, and we have checked that for $g_* = 4\pi$ most of the parameter space is excluded for $\Lambda \lesssim 10$ TeV, while for $g_* = 1$ the bound is relaxed to $\Lambda \lesssim 1$ TeV. Moreover, this operator induces severe constraints from dark matter scattering off nuclei for $m_{DM}\gtrsim 5$ GeV \cite{1411.3342}. 

\section{Conclusions}

As more and more parameter space is ruled out by experiments without any clear signal of dark matter discovery, it is timely to explore new venues and new regions of parameter space traditionally neglected. In this paper we have analyzed the case of dark matter with mass in the MeV range, {\it i.e.} below the reach of current direct detection experiments. This region is particularly interesting since can be probed at future direct detection experiments involving the dark matter scattering off electrons. Using a model independent approach, we have added to the standard model lagrangian all the dimension 6 effective operators that can involve dark matter, and we have properly taken into account the mixing between operators induced in the renormalization group running. Our main results are summarized in Figures \ref{fig:QU} and \ref{fig:LU}. As can be seen, large portions of parameter space are already probed in a model independent way. Although the exact value of the maximum scale $\Lambda$ already excluded highly depends on the structure and the size of the UV couplings, it is clear from the plots that, under our assumptions,  most of the parameter space to which future electron scattering and CMB experiments are sensitive is already ruled out. We stress that since most of the bounds involve scales below the top mass, in this case $\Lambda$ should be interpreted as the mass of some mediator generating the relevant operators. Our bound applies also to this case in the limit in which we neglect effects involving the resonant production of the mediator.

\acknowledgments

We would like to thank S. Bruggisser, S. Fichet, F. Iocco, B. Kavanagah, M. Taoso and A. Urbano for valuable discussions and suggestions. This work was supported by Funda\c{c}\~{a}o de Amparo \`{a} Pesquisa do Estado de S\~{a}o Paulo (FAPESP) and Conselho Nacional de Ci\^{e}ncia e Tecnologia (CNPq).

\appendix
\section{Useful Formulas}\label{app:formulas}

We present in this Appendix some useful formula. To set the notation, we write a generic coupling between the Dark Matter and the Standard Model fermions as
\be\label{app:lagr}
{\cal L} = \overline{\chi} \gamma^\mu \chi  \left[ \frac{c_{VVf} }{\Lambda^2} \overline{f}\gamma_\mu f + \frac{c_{VAf} }{\Lambda^2} \overline{f}\gamma_\mu \gamma_5 f \right] + \overline{\chi} \gamma^\mu \gamma_5 \chi  \left[ \frac{c_{AVf} }{\Lambda^2} \overline{f}\gamma_\mu f + \frac{c_{AAf} }{\Lambda^2} \overline{f}\gamma_\mu \gamma_5 f \right] .
\ee
Following Ref. \cite{1611.00368}, the lowest order chiral perturbation theory lagrangian coupling DM to mesons is given by
\be
{\cal L}_{\chi PT} = i\mbox{Tr} ([\partial_\mu\Pi,\Pi]\nu_{\chi}^{\mu})-\sqrt{2}f \mbox{Tr}(\partial^\mu \Pi a_\mu) , 
\ee
where as usual the mesons hermitian matrix reads
\be\label{app:mesmatrix}
\Pi=\left( 
\begin{matrix}
\frac{\pi_0}{\sqrt{2}} + \frac{\eta}{\sqrt{6}} & \pi^{+} & K^{+} \\
\pi^{-} & -\frac{\pi^{0}}{\sqrt{2}} + \frac{\eta}{\sqrt{6}} & K^{0} \\
K^{-} & \overline{K}^{0} & -2 \frac{\eta}{\sqrt{6}}
\end{matrix}
\right),
\ee
while the vector spurions including the DM currents are defined as
\be
\ba{rcl}
\nu_\chi^\mu &=& {\displaystyle \frac{\mbox{diag}(c_{VVu}, c_{VVd}, c_{VVs})}{\Lambda^2}\overline{\chi}\gamma^\mu\chi + \frac{\mbox{diag}(c_{AVu}, c_{AVd}, c_{AVs})}{\Lambda^2}\overline{\chi}\gamma^\mu\gamma^5\chi }, \\
a_\chi^\mu &=& {\displaystyle \frac{\mbox{diag}(c_{VAu}, c_{VAd}, c_{VAs})}{\Lambda^2}\overline{\chi}\gamma^\mu\chi + \frac{\mbox{diag}(c_{AAu}, c_{AAd}, c_{AAs})}{\Lambda^2}\overline{\chi}\gamma^\mu\gamma^5\chi .}
\ea
\ee
Using the previous definitions, the annihilation cross section $\overline{\chi}\chi \to \overline{M}M$ into mesons $M$, for a vector DM current, is given by
\begin{align}
\sigma v_{\rm rel} = \frac{c_{VM}^2\left(m_{DM}^2-m_{M}^2\right) \beta_M}{4 \pi  \Lambda^4} + \frac{v_{\rm rel}^2 c_{VM}^2\left(5 m_{M}^2 + 4m_{DM}^2\right) \beta_M}{96 \pi  \Lambda ^4},
\end{align}
where $m_M$ is the meson mass and the relevant couplings are
\be
c_{VK^0} = c_{VVd} - c_{VVs}, ~~ c_{VK^\pm} = c_{VVs} - c_{VVu}, ~~ c_{V\pi^\pm} = c_{VVd} - c_{VVu},
\ee
and 
\be
\beta_i=\sqrt{1-\frac{m_{i}^2}{m_{DM}^2}}.
\ee
For a vector-axial DM current we have instead that the annihilation cross section is p-wave, and it is given by
\begin{align}
\sigma v_{\rm rel} = \frac{v_{\rm rel}^2 c_{AM}^2 \beta_M \left(m_{DM}^2-m_{M}^2\right)}{24 \pi  \Lambda ^4},
\end{align}
where the relevant couplings are
\be
c_{AK^0} = c_{AVd} - c_{AVs}, ~~ c_{AK^\pm} = c_{AVs} - c_{AVu}, ~~ c_{A\pi^\pm} = c_{AVd} - c_{AVu} .
\ee

As explained in Section \ref{ssec:CMB}, the dominant contribution to the annihilation cross section is given by $\overline{\chi}\chi \to e^+ e^-$. Using the notation of Equation (\ref{app:lagr}), for a vector DM current we have
\begin{align}
	\sigma v_{\rm rel} &= \frac{\beta_e \bigg[c_{VVe}^2 \left(m_e^2+2 m_{DM
   }^2\right)-2 c_{VAe}^2 \left(m_e^2-m_{DM}^2\right)\bigg]}{2 \pi  \Lambda
   ^4}\\
   &- \frac{v_{\text{rel}}^2 \beta_e \bigg[c_{VAe}^2
   \left(2 m_e^2 m_{DM}^2-10 m_e^4+8 m_{DM}^4\right)+c_{VVe}^2 \left(-4
   m_e^2 m_{DM}^2+5 m_e^4+8 m_{DM}^4\right)\bigg]}{48 \pi  \Lambda ^4
   \left(m_e^2-m_{DM}^2\right)} \nonumber,
\end{align}
while for a vector-axial DM current the annihilation cross section is given by
\begin{eqnarray}
\sigma v_{rel} &=&  \frac{c_{AAe}^2 m_e^2 \beta_e}{2 \pi \Lambda ^4}\\
 &+&  \frac{v_{\text{rel}}^2 \beta_e \bigg[c_{AAe}^2 \left(22
 m_e^2 m_{DM}^2-17 m_e^4-8 m_{DM}^4\right)+4 c_{AVe}^2 \left(m_e^2 m_{DM}^2+m_e^4-2 m_{DM}^4\right)\bigg]}{48 \pi  \Lambda ^4 \left(m_e^2-m_{DM}^2\right)}.\nonumber
\end{eqnarray}

\bibliography{biblio}
\bibliographystyle{JHEP}

\end{document}